\documentclass[prb,aps,eqsecnum]{revtex4}
\usepackage{graphicx}
\newcommand{\al}{{\alpha}}

\newcommand{\e}{{\rm e}}
\newcommand{\ep}{\varepsilon}

\newcommand{\bdt}{{\bf T}}
\newcommand{\bdv}{{\bf V}}

\newcommand{\be}{\begin{equation}}
\newcommand{\ee}{\end{equation}}
\newcommand{\ba}{\begin{eqnarray}}
\newcommand{\ea}{\end{eqnarray}}

\newcommand{\nn}{\nonumber}
\newcommand{\la}{\label} 
\newcommand{\w}{\omega} 
\newcommand{\hT}{\hat{T}}
\newcommand{\hV}{\hat{V}}

\newcommand{\dt}{\Delta t}

\begin{document}
\title{Higher-order splitting algorithms for solving the nonlinear 
Schr\"odinger equation and their instabilities}

\author{Siu A. Chin}
\affiliation{Department of Physics, Texas A\&M University,
College Station, TX 77843, USA}

\begin{abstract}

Since the kinetic and the potential energy term of the
real time nonlinear Schr\"odinger equation can each be
solved exactly, the entire equation can be solved to any order 
via splitting algorithms. We verified the fourth
order convergence of some well known algorithms by solving
the Gross-Pitaevskii equation numerically.
All such splitting algorithms suffer from a latent 
numerical instability even when the total energy is very well conserved.
A detail error analysis reveals that the noise, or elementary
excitations of the nonlinear Schr\"odinger, obeys the Bogoliubov
spectrum and the instability is due to the exponential growth of high 
wave number noises caused by the splitting process.
For a continuum wave function, this instability is unavoidable
no matter how small the time step. For a discrete wave function,
the instability can be avoided only for $\dt\, k_{max}^2{<\atop\sim}2 \pi$,
where $k_{max}=\pi/\Delta x$.

\end{abstract}
\maketitle

\section {Introduction}

Taha and Ablowitz\cite{taha} have shown for 
some time that the first order pseudo-spectral, split-operator 
method is a very fast way of solving the nonlinear 
Schr\"odinger equation. Bandrauk and Shen\cite{band} later applied 
higher-order splitting algorithms with negative coefficients 
to solve the same equation. They regarded the nonlinear potential
as time-dependent. Since they can only estimate the intermediate-time nonlinear 
potential to second order, it is not proven that their higher-order algorithms
actually converge at fourth or sixth-order. Recently Javanainen and 
Ruostekoski\cite{java}
have shown by symbolic calculations that fourth-order algorithms are
possible by use of the ``latest" intermediate 
wave function in evaluating the nonlinear potential. Strauch\cite{stra},
by constructing a special operator that correctly propagates the nonlinear
potential term, proved that this use of the ``latest" intermediate
wave function is valid. 

This work shows that: 1) Javanainen and Ruostekoski's finding
is a direct consequence of Taha and Ablowitz' original work and a much 
simpler proof than that of Strauch is possible. 2) The
time-dependent potential method of Bandrauk-Shen and the time-independent
approach suggested by Javanainen and Ruostekoski both yielded identical 
second-order algorithms but different higher-order algorithms. 3) 
Verified numerically that algorithms derived by the time-independent method 
do converge to fourth-order when solving the Gross-Pitaevskii equation. 
4) All such splitting algorithms possess a latent numerical instability 
which causes the wave function to blow up despite excellent total energy
conservation. 5) The instability is shown to be due to the exponential growth of 
high wave number noises intrinsic to the splitting process. For a continuum wave 
function, this instability is unavoidable no matter how small is the time step. 
For a discrete wave function, this can only be avoided if $\dt{<\atop\sim}2\pi/k_{max}^2$, 
which forces $\dt$ to be very small if the discretization is very fine with 
a large $k_{mas}=\pi/\Delta x$. The next three sections summarize how higher order 
algorithms can be systematically derived and Section V discusses the instability in detail.

\section {Solving the nonlinear Schr\"odinger equation}

Consider the nonlinear Schr\"odinger equation defined by  
\be
i\frac{\partial \psi}{\partial t}=(-\frac12 \nabla^2 +g|\psi|^2)\psi.
\la{nl}
\ee 
The free particle propagation can be solved exactly
in operator form 
\be
\psi(\Delta t)=\e^{-i\Delta t \hT}\psi(0)
\la{kin}
\ee
where the operator $\hT=-\frac12 \nabla^2$. Since $\hT$ is diagonal in
k-space, (\ref{kin}) is usually solved by Fast Fourier Transforms (FFT).
Surprisingly, as shown by Taha and Ablowitz, the potential part 
of the equation
\be
i\frac{\partial \psi}{\partial t}=g|\psi|^2\psi.
\la{pot}
\ee
can also be solved exactly
\be
\psi(\Delta t)=\e^{-i\Delta t g|\psi(0)|^2}\psi(0).
\la{potsol}
\ee
This is because (\ref{pot}) exactly conserves $|\psi|^2$ (multiply (\ref{pot}) 
by $\psi^*$, the complex conjugated equation by $\psi$ and subtract) 
and the nonlinear potential is just a constant in (\ref{pot}). 
This is also clear from (\ref{potsol}), 
\be
|\psi(\Delta t)|^2=|\psi(0)|^2,
\ee
since $\psi(0)$ is only multiplied by a phase.
 Eq.(\ref{kin}) and (\ref{potsol}) are the basic
building blocks for constructing splitting algorithms for solving the
nonlinear Schr\"odinger equation. Eq.(\ref{potsol}) is the 
fundamental justification for using the ``latest" wave function in computing 
the nonlinear potential\cite{java}. (See also below).
Define a {\it time-independent} operator $\hV$ such that
\be
\hV|\psi(t)\rangle=g|\psi(t)|^2|\psi(t)\rangle.
\ee
Note that $\hV$ only acts on $|\psi(t)\rangle$ and does not act on 
its own eigenvalue $g|\psi(t)|^2$. It follows that
\be
\e^{-i\Delta t \hV}|\psi(t)\rangle=\e^{-i\Delta t g|\psi(t)|^2}|\psi(t)\rangle.
\la{potop}
\ee 
The crucial point here is that $\hV$ has no time-dependence, when
it acts on any $|\psi(t)\rangle$, it produce the  
eigenvalue $g|\psi(t)|^2$. 
The resulting time-dependence of the nonlinear potential is due entirely 
to the state vector $|\psi(t)\rangle$ and not to the operator $\hV$. 
The exact solution can then be written in operator form as
\be
|\psi(t)\rangle=\e^{-it(\hT+\hV)}|\psi(0)\rangle.
\la{exsol}
\ee
For our purpose here, we only need to know (\ref{potop}) and not
the explicit form of $\hV$. For an elegant, but rather abstract
construction of $\hV$, see Strauch's\cite{stra} recent work. 

\section {Deriving Splitting algorithms}

To solve (\ref{exsol}) by splitting algorithms,
one factorizes the evolution operator to any order 
with a suitable set of coefficients $\{t_i,v_i\}$ via
\be
\e^{\ep (\hT+\hV)}=\prod_{i}\e^{\ep t_i\hT}\e^{\ep v_i \hV},
\ee 
where we have denoted $\ep=-i\Delta t$. For example, we can have
the second order algorithm 2A as 
\be
\psi(\Delta t)=\e^{\frac12 \ep \hV}
                 \e^{\ep \hT}
			 \e^{\frac12 \ep \hV}\psi(0)=\e^{\frac12 \ep g|\phi|^2}
                 \e^{\ep \hT}
			 \e^{\frac12 \ep g|\psi(0)|^2}\psi(0)
\la{alg2a}
\ee
where according to (\ref{potsol}) or (\ref{potop}), we must take
\be
\phi=\e^{\ep \hT}
			 \e^{\frac12\ep g|\psi(0)|^2}\psi(0).
\ee
Algorithm 2A only requires one-pair of FFT (forward and backward) to
achieve second-order accuracy, which is the same number of FFT needed
for a first-order algorithm.
If the nonlinear potential is treated as a time-dependent potential, 
as done by Bandrauk and Shen\cite{band}, 
then we would have the algorithm\cite{suzu93,chinc02}
\be
\psi(\Delta t)=
{\rm e}^{-i\frac12 \Delta t V(\Delta t)}
{\rm e}^{-i\Delta t \hT}
{\rm e}^{-i\frac12 \Delta t V(0)}\psi(0).
\la{t2a}
\ee
In this case, since the last factor is only
a phase,
\be
V(\Delta t)=g|\psi(\Delta t)|^2=g|\phi|^2 ,
\ee
the result is the same as (\ref{alg2a}). If one ignores the
time-dependence\cite{adh} and uses
$V(\Delta t)=V(0)=g|\psi(0)|^2$, then algorithm (\ref{t2a}) is
degraded to first order. 

Similarly one has the second-order algorithm 2B, 
\be
\psi(\Delta t)=\e^{\frac12\ep \hT}
				 \e^{\ep \hV}
				 \e^{\frac12\ep \hT}\psi(0)=\e^{\frac12\ep \hT}
				 \e^{\ep g|\phi|^2}
				 \e^{\frac12\ep \hT}\psi(0),
\la{alg2b}
\ee
where here
 \be
\phi=\e^{\frac12 \ep \hT}\psi(0).
\ee
In the time-dependent potential approach, one would have instead,
\be
\psi(\Delta t)=
{\rm e}^{-i\frac12 \Delta t \hT}
{\rm e}^{-i{\Delta t} V(\Delta t/2)}
{\rm e}^{-i\frac12 \Delta t \hT}
\psi(0).
\ee
One must now evaluate $V(\Delta t/2)=g|\psi(\Delta t/2)|^2$.
Since the algorithm is only second order, one can simply approximate
the midpoint wave function to first order,
\be
\psi(\Delta t/2)=
{\rm e}^{-i{\frac12 \Delta t} V(\Delta t/2)}
{\rm e}^{-i\frac12 \Delta t \hT}
\psi(0),
\ee
and therefore
\be
|\psi(\Delta t/2)|^2=
|{\rm e}^{-i\frac12 \Delta t \hT}\psi(0)|^2.
\ee
Again, the result is the same as (\ref{alg2b})

For fourth and higher order algorithms, the time-dependent potential 
approach cannot be easily implemented. 
It is much more efficient to use the ``latest" intermediate wave function
than to estimate the intermediate-time wave function to third or higher order.
Thus higher order algorithms are currently possible only with the use of 
the time-independent formalism based on the original finding of
Taha and Ablowitz.

The fourth-order Forest-Ruth (FR) \cite{fr90} algorithm, which 
is the triplet concatenation\cite{cre89,suzu90,yos90} of algorithm 2A 
\be
{\cal T}_{FR}(\ep)=	{\cal T}_{2A}(c_1\ep)
{\cal T}_{2A}(c_0\ep)
{\cal T}_{2A}(c_1\ep)
\la{fr}
\ee
with $c_1=1/(2-2^{1/3})$ and $c_0=-2^{1/3}/(2-2^{1/3})$ 
has been verified by Javanainen and Ruostekoski as obeying the
``latest" intermediate wave function rule. However, since this triplet 
concatenation will convert any second-order split algorithm to 
fourth-order, verifying this algorithm alone does not constitute a check
on more general fourth-order algorithms. (Recall that
algorithm 2A can also be derived from the time-dependent approach
without explicitly invoking the ``latest" wave function rule.) 
(Javanainen and Ruostekoski have also verified the ``latest" wave function  
rule on a class of third-order algorithms independent of 2A.)
To seal this
loop-hole in our verification process, we also consider more general 
fourth-order algorithms 
previously studied by McLachlan\cite{mcl95} with 9 operators,
\be {\cal T}_{M}= \dots 
\exp(\ep t_0 \hV) 
\exp(\ep v_1 \hT) 
\exp(\ep t_1 \hV) 
\exp(\ep v_2 \hT) 
\exp(\ep t_2 \hV). 
\la{algm1}
\ee
The factorization is left-right symmetric and only operators
from the center to the right are indicated.	The fourth-order
order condition requires\cite{chin972} that
\be
v_1=\frac12-v_2,\quad
t_2=\frac16-4 t_1 v_1^2, \quad
t_0=1-2(t_1+t_2),
\ee
\be
w=\sqrt{3-12t_1+9 t_1^2},\quad
v_2=\frac14 \left(1\mp\sqrt{\frac{9 t_1-4 \pm 2w}{3t_1}}\,\,\right)
\ee
and that the free parameter $t_1<0$. This algorithm requires 4 pairs of
FFT but has a much smaller energy error and greater stability
than that of FR. (The coefficient designation does
not match the the operators because the algorithm has been adapted 
from its classical version by interchanging $\hT\leftrightarrow \hV$.)
There are four solution branches for $v_2$.
The choice of 
$$t_1=\frac{121}{3924}(12-\sqrt{471})\approx -0.299$$
with 
\be
v_2=\frac14 \left(1+\sqrt{\frac{9 t_1-4 + 2w}{3t_1}}\,\,\right)
\la{v2m1}
\ee
reproduces McLachlan's\cite{mcl95} recommended algorithm. 
By varying $t_1$ and using different branches of $v_2$, it is
possible to optimize the algorithm for specific applications.  
For application in the next section, the results are not very sensitive
to the branch of $v_2$ nor the
choice of $t_1$, as long as $t_1$ is in the range of [-0.1, -0.4].
More higher-order splitting algorithms can be found in Refs.\cite{hairer02,mcl02,bm02,lr04}.

\section {Numerical verifications}

To verify the order of convergence of these algorithms, we apply them 
to the Gross-Pitaevskii equation with a harmonic trap in 1D,
\be
i\frac{\partial \psi}{\partial t}
=(-\frac12 \frac{d^2}{dx^2}+\frac12\omega^2 x^2 +g|\psi|^2)\psi.
\ee
To gauge the accuracy of any algorithm, we monitor the fluctuation of
the total $E$, 
\be
E=\int_{-\infty}^\infty dx\, \psi^*(t)
( -\frac12 \frac{d^2}{dx^2}+\frac12\omega^2 x^2+\frac12g|\psi(t)|^2)\psi(t)
\la{eng} 
\ee
If the time evolution is exact, $E$ would remain a constant.
For $\omega=1$, $g=5$ and $\psi(0)=\psi_0(x)$, 
the ground state wave function of the harmonic trap, the initital total energy is
\be
E=\frac12+\frac5{2\sqrt{2\pi}}\approx 1.497355701.
\ee
The $x$-interval used is [-20:20] with $2^9=512$ grid-points.
The results are unchanged if one doubles the grid-points. 
In Fig.1 we plot $E$ as a function of time for algorithm 2A at 
$\Delta t=0.05$ and $\Delta t=0.025\,$. One observes that the
energy fluctuation at $\Delta t=0.025$ is about 1/4 of that at
$\Delta t=0.05$, as befitting a second order algorithm. The results for
fourth-order algorithms FR (Forest-Ruth) and M (McLachlan) 
at $\Delta t=0.05$ are also shown. It
is clear that even if one take 1/4 of algorithm 2A's error at $\Delta t=0.025$,
corresponding to $\Delta t=0.0125$, that error is still much larger than
those of fourth-order algorithm FR and M ({\it i.e.}, running algorithm
2A four times at $\Delta t=0.0125$, using 4 pairs of FFT, would
still be inferior to algorithm FR which uses only 3 pairs of FFT). 

In Fig.2 we greatly magnified the scale so that the fluctuations in
the fourth-order algorithms are also visible. This time, when the step size 
of algorithm FR is half, the error in $E$ is reduced by a factor of
16, confirming the fourth-order convergence of the algorithm. The energy error of
algorithm M at $\Delta t=0.025$ is $\approx 10^{-6}$, which is too small for
a visual comparison.
  
In both Figures 1 and 2, the total energy eventually blows up for all calculations,
despite the fact that total energy error is only $10^{-6}$ for McLachlan's
algorithm. This instability is directly related to the strength of the 
nonlinear potential. The rather large value of $g=5$ was chosen so that the 
instability would show up after a short run. The energy blow up can be delayed,
but not eliminated, by reducing $\dt$. (See further discussion in Section VI.)

\section {The cause of instability}

The eventual instability as shown in Figures 1 and 2 demands an understanding
of its fundamental cause. To study this, we decompose the general wave function into
Fourier components and focus on the
propagation of a single component with wave vector $p$ in 1D,
\be
\psi(x,t)=A\e^{i p x-i\w t}.
\la{pw}
\ee
This is a solution to (\ref{nl}) if $\w$ 
is given by
\be
\w=\frac12 p^2+g|A|^2=E_p+U,
\ee
where we have denoted $E_p=\frac12 p^2$ and $U=g|A|^2$.  
Suppose now the spatial part of $\psi$ is contaiminated, due to numerical errors,
by very small amplitude, side-band wave vectors $p+k$ and $p-k$
so that
\be
\psi(x)=A\e^{i p x}+a\e^{i(p+k)x}+b\e^{i(p-k)x},
\la{psq}
\ee
how will the error amplitudes $a$ and $b$ be propagated by splitting algorithms?
(This side-band analysis was inspired by the classical work on Fourier
analysis of nonlinearly interacting waves\cite{whit}.) 
The effect of $\e^{-i\dt\hT}$ on $\psi(x)$ is trivial; all amplitudes are
multiplied by a phase,
\ba
A^\prime&=&\e^{-i\dt E_p}A\nn\\
a^\prime&=&\e^{-i\dt E_{p+k}}a\la{et}\\
b^\prime&=&\e^{-i\dt E_{p-k}}b\nn.
\ea
To compute $\e^{-i\dt\hV}\psi(x)$, one must compute $|\psi(x)|^2$ using (\ref{psq}).
The result, by keeping terms only to first order in
$a$ and $b$, is
\ba
A^\prime&=&\e^{-i\dt U}A\nn\\
a^\prime&=&\e^{-i\dt U}[a-i\dt(U a+gA^2b^*)]\la{ev}\\
b^\prime&=&\e^{-i\dt U}[b-i\dt(U b+gA^2a^*)]\nn.
\ea
Thus the first order splitting algorithm $\e^{-i\dt\hV}$$\e^{-i\dt\hT}\psi(x)$
modifies the amplitudes by composing (\ref{et}) with (\ref{ev}), yielding
\ba
A_{n+1}&=&\e^{-i\dt(E_p+ U)}A_n\la{oneA}\\
a_{n+1}&=&\e^{-i\dt(E_{p+k}-E_k+U)}
[a_n\e^{-i\dt E_k}-i\dt U(a_n\e^{-i\dt E_k}
+(b_n\e^{-i\dt E_k})^*\e^{-i2\delta_n})]\la{onea}\\
b_{n+1}&=&\e^{-i\dt(E_{p-k}-E_k+U)}
[b_n\e^{-i\dt E_k}-i\dt U(b_n\e^{-i\dt E_k}
+(a_n\e^{-i\dt E_k})^*\e^{-i2\delta_n})]\la{oneb},
\ea
where we have defined
\be
A_n=|A_n|\e^{-i\delta_n}.
\ee
The algorithm correctly propagates $A$ and preserves the norm $|A|$,
\be
A_n=\e^{-in\dt(E_p+ U)}A_0.
\ee 
For notational clarity, we will take $A_0$ to be real with
$\delta_0=0$ so that we don't have to keep track of this initial
phase, yielding 
\be
\delta_n=n\dt(E_p+U).
\ee
(Keeping the initial phase simply transfers it to subsequent
amplitudes and has no bearing on the issue of instability.)  
To see the growth in $a$ and $b$, we factor out their overall phases as follow
\ba
a_n&=&\e^{-in\dt(E_{p+k}-E_k+U)}\alpha_n,\nn\\ 
b_n&=&\e^{-in\dt(E_{p-k}-E_k+U)}\beta_n,
\ea 
and reduce (\ref{onea}) and (\ref{oneb}) to
\ba
\alpha_{n+1}&=&\alpha_n\e^{-i\dt E_k}-i\dt U(\alpha_n\e^{-i\dt E_k}
+(\beta_n\e^{-i\dt E_k})^*)\la{ga}\\
\beta_{n+1}&=&\beta_n\e^{-i\dt E_k}-i\dt U(\beta_n\e^{-i\dt E_k}
+(\alpha_n\e^{-i\dt E_k})^*)\la{gb}.
\ea
These two equations can also be interpreted as a first-order
splitting algorithm,
with the ``kinetic" term giving
\ba
\alpha^\prime &=&\e^{-i\dt E_k}\alpha\nn\\  
\beta^\prime &=&\e^{-i\dt E_k}\beta
\la{kterm} 
\ea
and the ``potential" term producing 
\ba
\alpha^\prime &=&\alpha-i\dt U(\alpha+\beta^*) \nn\\  
\beta^\prime &=&\beta-i\dt U(\beta+\alpha^*). 
\la{vterm}
\ea
A closer examination reveals that (\ref{kterm})
and (\ref{vterm}) are exact solutions to following equations
\be
i\frac{d\alpha}{dt}=E_k\alpha,\quad	i\frac{d\beta}{dt}=E_k\beta,
\ee
\be
i\frac{d\alpha}{dt}=U(\alpha+\beta^*),\quad	i\frac{d\beta}{dt}=U(\beta+\alpha^*).
\ee
Thus the algorithm is trying to solve the original unsplitted equations
\ba
i\frac{d\alpha}{dt}=(E_k+U)\alpha+U\beta^*\nn\\
i\frac{d\beta}{dt}=(E_k+U)\beta+U\alpha^*,
\la{beq}
\ea
which have general solutions of the form 
\be
\alpha=c\e^{-i\Omega_k t}+d\e^{i\Omega_k t},
\la{amp}
\ee
with
\be
\Omega_k
=\sqrt{E_k(E_k+2U)}.
\la{bspec}
\ee
This is the famous Bogoliubov spectrum\cite{bog} of elementary excitations in
a uniform Bose gas. It shows up here because the nonlinear Schr\"odinger
equation is just the Gross-Pitaevskii equation for describing a uniform
Bose-Einstein condensate\cite{pet}. The Bogoliubov
spectrum in the current context, is the background ``noise" excitations of 
the nonlinear Schr\"odinger equation.
If one were able to solve (\ref{beq}) exactly via (\ref{amp}), there would
be no instability because the amplitude of $\alpha$ in (\ref{amp}) is finite. However,
when (\ref{beq}) is solved by splitting, (\ref{vterm}) no longer preserves the 
norm and the modulus of these error terms at selected ranges of $k$ will 
grow exponentially. 

To study this growth, take
$\beta_0=\alpha_0$, so that the splitting forms (\ref{kterm})
(\ref{vterm}) simplify to  
\ba
\alpha^\prime &=&\e^{-i\dt E_k}\alpha\la{alspa}\\  
\alpha^\prime &=&\alpha-i\dt U(\alpha+\alpha^*).
\la{alspb}
\ea
Now we assert without giving a detail proof that beyond first-order, for 
any splitting algorithm in solving the nonlinear Schr\"odinger equation,
the error Fourier components will grow correspondingly according to splitting 
(\ref{alspa}-\ref{alspb})
with the same splitting coefficients. For example, corresponding to algorithm 2A,
the growth of the error Fourier components is given by 
\ba
\alpha_1 &=&\alpha_0-i\frac12 \dt U(\alpha_0+\alpha_0^*)\nn\\
\alpha_2 &=&\e^{-i\dt E_k}\alpha_1\la{eal2a}\\  
\alpha_3 &=&\alpha_2-i\frac12 \dt U(\alpha_2+\alpha_2^*)\nn
\ea
The subscripts here simply label the individual steps in the algorithm.
The last labelled value is the updated variable after one time step. Denoting
this updating as ${\cal E}_{2A}(\dt)$, the error growth of the Forest-Ruth algorithm
is then 
\be
 {\cal E}_{FR}(\dt)={\cal E}_{2A}(c_1\dt)
 {\cal E}_{2A}(c_0\dt)
 {\cal E}_{2A}(c_1\dt)
 \la{ealfr}
\ee 
and McLachlan's algorithm as
\ba
\alpha_1 &=&\alpha_0-i(t_2 \dt) U(\alpha_0+\alpha_0^*)\nn\\
\alpha_2 &=&\e^{-i(v_2\dt) E_k}\alpha_1\nn\\  
\alpha_3 &=&\alpha_2-i(t_1 \dt) U(\alpha_2+\alpha_2^*)\nn\\
\alpha_4 &=&\e^{-iv_1\dt E_k}\alpha_3\nn\\  
&&\dots\dots, \quad{\rm etc.}
\la{ealmcl}
\ea
To verify the validity of our assertion, we run the normal algorithm on
an initial wave function having the $p=0$ component with amplitude $A=1$, and
all other Fourier components set to $\e^{-25}$, at $g=5$ and $\dt=0.2$. 
The resulting Fourier amplitudes are then outputted every 
time steps for seven time steps. Their modulus are shown as plus signs for 
the above three algorithms in Figs. 3-5. Instead of plotting the magnitude
of these Fourier amplitudes as a function of $k$, we plot them as a function of
$\dt E_k/\pi$, which is more revealing. Also plotted as solid lines,
are the predicted error amplitudes given by (\ref{eal2a}), (\ref{ealfr})
and (\ref{ealmcl}) for seven time steps. The perfect agreement in all three
cases confirms our assertion and our side-band analysis.

To understand the pattern of instability as shown in Figs. 3-5, we
rewrite the splitting forms (\ref{alspa}) and (\ref{alspb}) as matrices 
acting on the real and imaginary part of $\alpha$
\begin{equation}
\left(\begin{array}{c}
       \al^\prime_R\\
       \al^\prime_I
      \end{array}\right)
= 
{\bf T}(\dt)
\left(\begin{array}{c}
       \al_R\\
       \al_I
      \end{array}\right), \quad
\left(\begin{array}{c}
       \al^\prime_R\\
       \al^\prime_I
      \end{array}\right)
= 
{\bf V}(\dt)
\left(\begin{array}{c}
       \al_R\\
       \al_I
      \end{array}\right),
\label{tveq}
\end{equation}
with
\be
{\bf T}(\dt)=	
\left(\begin{array}{cc}
                 c & s \\
                 -s&  c
      \end{array}\right),\quad
{\bf V}(\dt)= 
\left(\begin{array}{cc}
                 1 & 0 \\
                 -2u & 1
      \end{array}\right),
\label{tvm}
\ee
and where we have defined 
\be
c=\cos(x), \quad s=\sin(x), \quad x=\dt E_k, \quad{\rm and} \quad u=\dt U.
\ee
The updating matrix corresponding to algorithm 2A is therefore
\ba
{\bf M}_{2A}(\dt)&=&\bdv(\frac12 \dt)\bdt(\dt)\bdv(\frac12 \dt)\nn\\
&=&\left(\begin{array}{cc}
                 c-us & s \\
             (u^2-1)s-2uc\quad& c-us
      \end{array}\right).
\label{m2a}
\ea
This is a special form of a matrix with equal diagonal elements
and unit determinant. This is due to the left-right
symmetric form of the matrix product ( {\it i.e.}, the algorithm
is time-reversible\cite{chin053}) and that both $\bdt$ and $\bdv$
have unit determinant. Such a matrix has the special property
that its eigenvalue is given by
\be
e_{1,2}=C\pm\sqrt{C^2-1},
\ee  
where $C$ is just the diagonal element (or half of the trace of the matrix).
If $|C|<1$, the
eigenvalues are complex with unit modulus and the algorithm is
stable. If $|C|>1$, the eigenvalues are real
with one eigenvalue always greater than unity. Thus by just plotting 
$C$ against $x=\dt E_k$, one can immediately determine the regions of
instability. For algorithm 2A, we have
\be
C(x)=\cos(x)-u\sin(x)=C_0\cos(x+\delta).
\la{c2a}
\ee
with
\be
C_0=\sqrt{1+u^2}\quad {\rm and}\quad \delta=\tan^{-1}u.
\ee
It is then immediately clear that as long as $u\neq 0$, the algorithm is 
unstable for $x$ in the interval
$[n\pi-2\delta,n\pi]$ where $n=1,2,3\dots$ etc..
At a fixed $U$, decreasing $\dt$ reduces $u$ and $\delta$, and hence 
the width of the instability region, but does not remove the
instability (but see further discussion in the next section).
In Fig.3, this $C$-function is plotted and lowered to -28 so that
the interval where $|C(x)|>1$ can be directly compared with the observed 
regions of instability. The peak instability
occurs at $x=n\pi-\delta$ with the maximum eigenvalue
\be
|e_{1,2}|= \sqrt{1+u^2}+\sqrt{u}.
\ee
For $\dt=0.2$ and $U=5$, we have $u=1$, 
$\delta=\pi/4$ and $|e|=1+\sqrt{2}$. After seven iterations, the e-fold
increase of the peaks would be $\log((1+\sqrt{2})^7)=6.16962$, which is the 
six e-fold increase of amplitude observed in Fig.3. Thus we have completely
accounted for, both qualitatively and quantitatively, the pattern of instability
as shown in Fig.3. The corresponding $C$-functions
for the Forest-Ruth and the McLachlan algorithm are also plotted
in Fig.4 and 5. Their $C$-functions
are too lengthy for a written display. (The analytical expression for 
McLachlan's $C$-function is more than a page long using Mathematica.)

By comparing Fig.3 and 4, one sees that the Forest-Ruth algorithm has a greater 
error growing rate than 2A. We will see in the next section
that this is precisely the reason why  
the FR algorithm blew up earlier than 2A in Fig.1.
Finally, as shown in Fig.5, McLachlan's algorithm manages to shift the $C$-function 
is such a way that the error peaks at $x/\pi=1,3$ are nearly 
eliminated. 

Further insights into the origin of this instability can be gained
by representing ${\bf T}(\dt)$ and ${\bf V}(\dt)$ in terms of traceless
matrices,
\be
{\bf T}(\dt)=	
\exp\left[\dt\left(\begin{array}{cc}
                 0 & E_k \\
                 -E_k&  0
      \end{array}\right)\right], \quad
{\bf V}(\dt)=\exp\left[\dt
\left(\begin{array}{cc}
                 0 & 0 \\
                 -2U & 0
      \end{array}\right)\right].
\label{exptv}
\ee
One can then immediately identify the unsplitted evolution operator as
\be	
\exp\left[\dt\left(\begin{array}{cc}
                 0 & E_k \\
                 -E_k-2U&  0
      \end{array}\right)\right]=\left(\begin{array}{cc}
                 \cos(\Omega_k\dt) & \sin(\Omega_k\dt) \\
                 -\sin(\Omega_k\dt)&  \cos(\Omega_k\dt)
      \end{array}\right),
\ee
which is that of a harmonic oscillator with the Bogoliubov spectrum $\Omega_k$.
Were one able to split it alternatively as
\be
{\bf T}^\prime(\dt)=	
\exp\left[\dt\left(\begin{array}{cc}
                 0 & E_k \\
                 0&  0
      \end{array}\right)\right], \quad
{\bf V}^\prime(\dt)=\exp\left[\dt
\left(\begin{array}{cc}
                 0 & 0 \\
                 -E_k-2U & 0
      \end{array}\right)\right],
\label{exptvp}
\ee
one would recover the stability criterion normally associated with
the harmonic oscillator. For example, the corresponding second-order algorithm
2A, $\bdv^\prime(\frac12 \dt)\bdt^\prime(\dt)\bdv^\prime(\frac12 \dt)$,
would then yield a $C$-function of
\be
C= 1-\frac12 \Omega_k^2\dt^2,
\ee
which limits stability to $\dt\leq 2/\Omega_k$, a well known
result. This limit is actually worse than $x\leq\pi-2\delta$,
which, as $U\rightarrow 0$, is $\dt\leq\pi/E_k$.
Our original splitting (\ref{exptv}) is therefore better
the usual harmonic oscillator splitting (\ref{exptvp}). Moreover, in contrast
to Fig.3, the usual harmonic oscillator splitting would have {\it no} stable
region whatsoever beyond $\dt E_k{>\atop\sim}\pi$!

In this section we have shown that the error growing pattern of any splitting
algorithms when solving the nonlinear Schr\"odinger can be analytically
understood. The instability is due to the exponential amplification of 
high $k$ noises at
$E_k{>\atop\sim}\pi/\dt$. 

\section {The instability of the Gross-Pitaevskii wave function}

We now repeat the calculations of Fig.1 at $\dt=0.05$ for 1200 time steps, to
the point where the algorithm FR begins to blow up. We plot in Figs. 6-8, the
modulus of the $k$-space wave function $|\psi(k)|$	as a function of $\dt E_k/\pi$
at every 100th time step. The initial Gaussian wave function is the straightline
seen plunging down close to vertical axis. Because of limited numerical precision,
that line levels off to some random values 
around $\e^{-35}\approx 10^{-16}$ at high $E_k$. 
These are the initial random errors of the wave function. When the algorithm
acts on the wave funtion, these random errors
are amplified successively and grow in time. For algorithm 2A, Fig.6 shows
error peaks at $x/\pi=1,2$ and 4, which is in agreemwnt with Fig. 3, but
no discernable peak is seen near $x/\pi=3$. For the Forest-Ruth algorithm, 
Fig.7 shows a promenient peak at $x/\pi=1$, followed by a peak-shoulder structure
at $x/\pi=2$ and 4, in agreement with Fig.4.
For McLachlan's agorithm, Fig.8 shows that the error peak at $x/\pi=1$ is 
conspicuously absent, and only peaks at $x/\pi=2,4$ are visible. This is in
excellent agreement with the predicted error structure of Fig.5. In the case of
the Forest-Ruth algorithm, the error peak at $x/\pi=1$ has grown sufficiently
to distort the wave function and cause the energy to blow up. These 
exponentially growing error peaks are like ticking time bombs, harmless at first,
but eventually overwhelm and destroy the wave function.

For a continuum wave function, this instability is unavoidable as long
as $\dt$ is finite. However, for a discrete wave function defined at
only $N$ grid points, there is a loop-hole. 
For a finite $N$-point calculation, the maximum $k$ vector is
$k_{max}=N \pi/L$ so that
$\dt E_k/\pi$ extends only out to $(0.05)0.5(512\pi/40)^2/\pi\approx 12.9$,
as shown in Figs.6-8. Thus one can take advantage of this and force stability by 
making $\dt$ so small that  
\be
\dt E_k^{max}<x_{min},
\la{cri}
\ee
where $x_{min}$ is the smallest value of $x$ such that $|C(x)|=1$ and 
$E_k^{max}=\frac12 k_{max}^2$. For most algorithms at small $\dt$,
$x_{min}\approx\pi$. This criterion (\ref{cri}) simply
shrinks the entire range of $E_k$ values to below the first instability point.
Thus the RF calculation would be stable for $\dt<\pi/(0.5(512\pi/40)^2)=0.0039$.
A more refined calculation at higher $N$ would required an even smaller $\dt$.
Such as small $\dt$ would make long-time simulation very time consuming.     
On the other hand, (\ref{cri}) also 
implies that stability can be achieved by lowering $k_{max}$, {\it i.e.}, using
fewer grid points. For example, at $N=128$, $\pi/(0.5(128\pi/40)^2)=0.062$.
When the FR algorithm is rerun at $\dt=0.05$ but with $N=128$,  
the total energy is indeed stable out to $t=300$. However, the wave function now looked
very jagged. Thus for long time simulations, one muct choose $\dt$ and $N$ judiciously.      

The instability observed here is very similar to the ``resonance"
instability of multiple-time step algorithms used in 
biomolecular simulations\cite{chin042}. There,  stability requires
that $\dt<\pi/\omega$, where $\omega$ is the faster physical frequency
in the problem. The latency in the energy blow-up has also been observed in density
functional calculations using split algorithms\cite{sug}. The energy
blow-up there is more gradual, but it is undoubtedly related to the nonlinear
Kohn-Sham density used, for which the nonlinear Schr\"odinger equation is
the simplest prototype.

\section {Conclusions}

In this work we have shown how splitting algorithms of any order
can be devised to solve the nonlinear Schr\"odinger equation. The key
ingredient is the exact solution of the potential equation (\ref{potsol}),
as pointed out earlier by Taha and Ablowitz\cite{taha}. This explains 
Javanainen and Ruostekoski's finding\cite{java} without the need to
construct Strauch's special operator\cite{stra}. Solution
(\ref{potsol}) clearly generalize to the case where
$g|\psi|^2\rightarrow v(|\psi|)$, implying that this class of general 
nonlinear equations can also be solved by splitting algorithms. 

In the course of verifying these alogrithms by solving the 
Gross-Pitaevskii equation, a latent instability is observed in 
all the algorithms. This instability persists regardless of the
order of the algorithm and despite excellent total energy
conservation. A detail error analysis reveals that this instability
is intrinsic to splitting algorithms and can only be avoided
if (\ref{cri}) is satisfied. 

The main advantage of higher-order algorithms is that a larger $\dt$ can 
be used for more efficient simulations. However the stability criterion (\ref{cri})
dictates a small $\dt$ regardless of order, thus negating much
of the presumed advantage of using higher order algorithms. (Of course,
higher order algorithm are useful for short time simulations,
where results can be obtained prior to the blow-up.) This work
also suggests that one must not use just any higher order algorithm, 
such FR, but higher order algorithm with a higher $x_{min}$, such as
McLachlan's algorithm. How algorithms can be derived systematically
with a higher $x_{min}$ is a fitting subject for a future study.

\begin{acknowledgments}
I thank E. Krotscheck for many valuable discussions over the years
on this subject. 
\end{acknowledgments}
\vspace{.5 truein}

\newpage
\centerline{REFERENCES}

\newpage
\begin{figure}
	\vspace{0.5truein}
	\centerline{\includegraphics[width=0.8\linewidth]{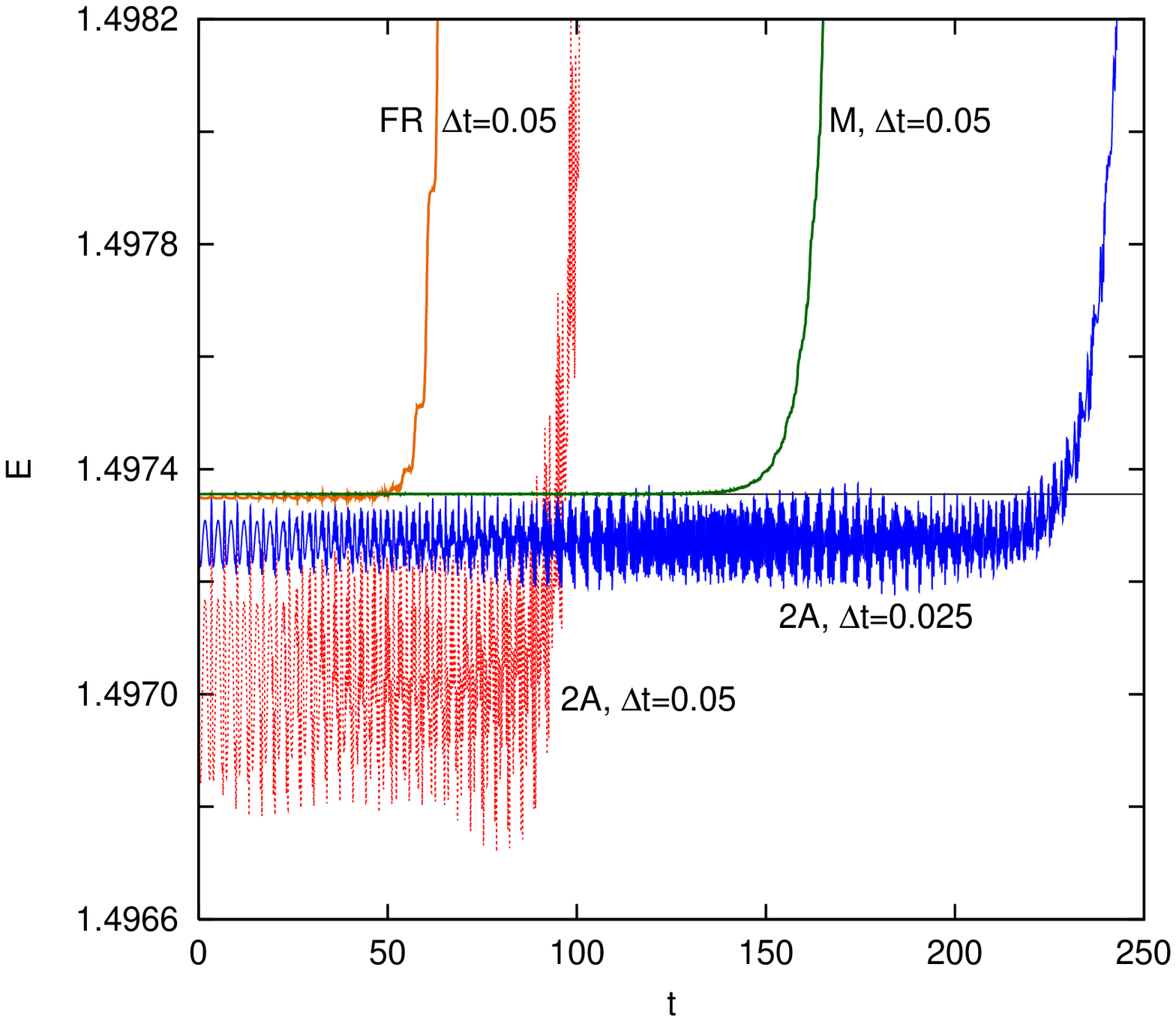}}
	\vspace{0.5truein}
\caption{(Color online) The fluctuation in the total energy E (\ref{eng}),
when solving the real time Gross-Pitaevskii equation by 
second order algorithm 2A and fourth-order algorithms FR (Forest-Ruth)
and M (McLachhan).
\label{fig1}}
\end{figure}

\newpage
\begin{figure}
	\vspace{0.5truein}
	\centerline{\includegraphics[width=0.8\linewidth]{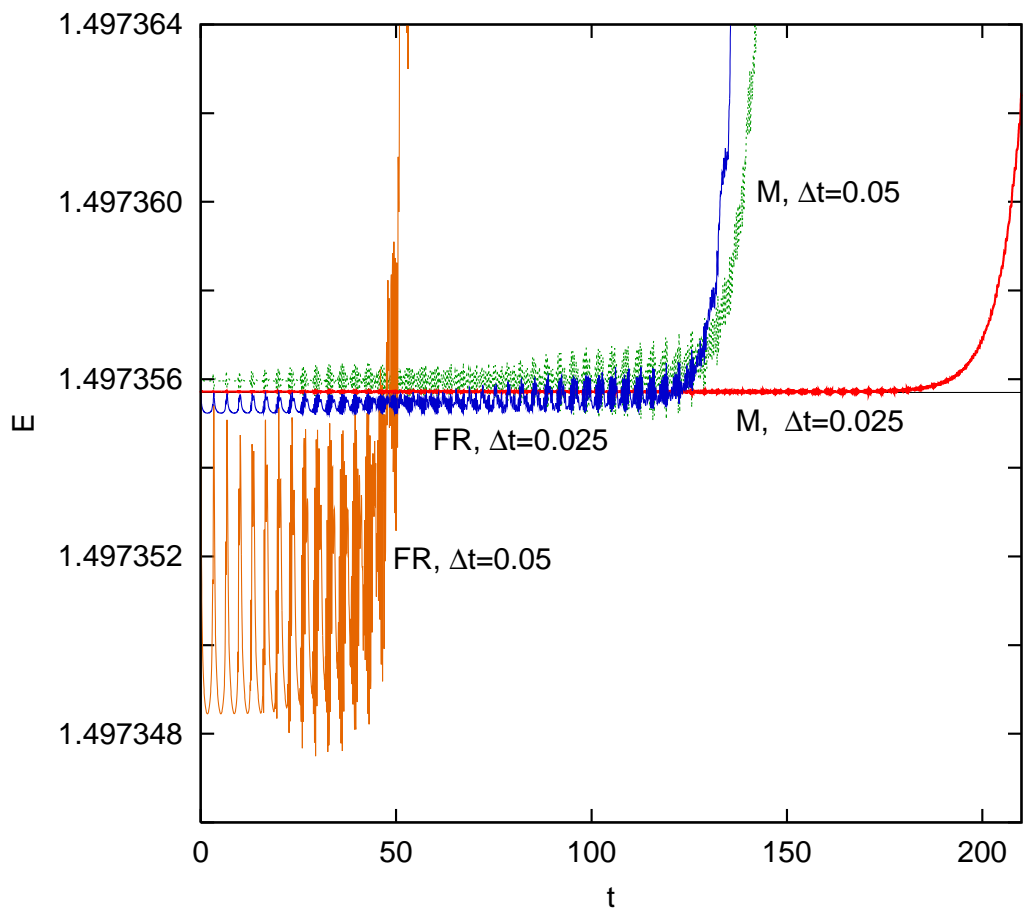}}
	\vspace{0.5truein}
\caption{(Color online) A magnified view of the fluctuation in the total energy
 of two fourth-order algorithms FR
and M at two time-step sizes.
\label{fig2}}
\end{figure}

\newpage
\begin{figure}
	\vspace{0.5truein}
	\centerline{\includegraphics[width=0.8\linewidth]{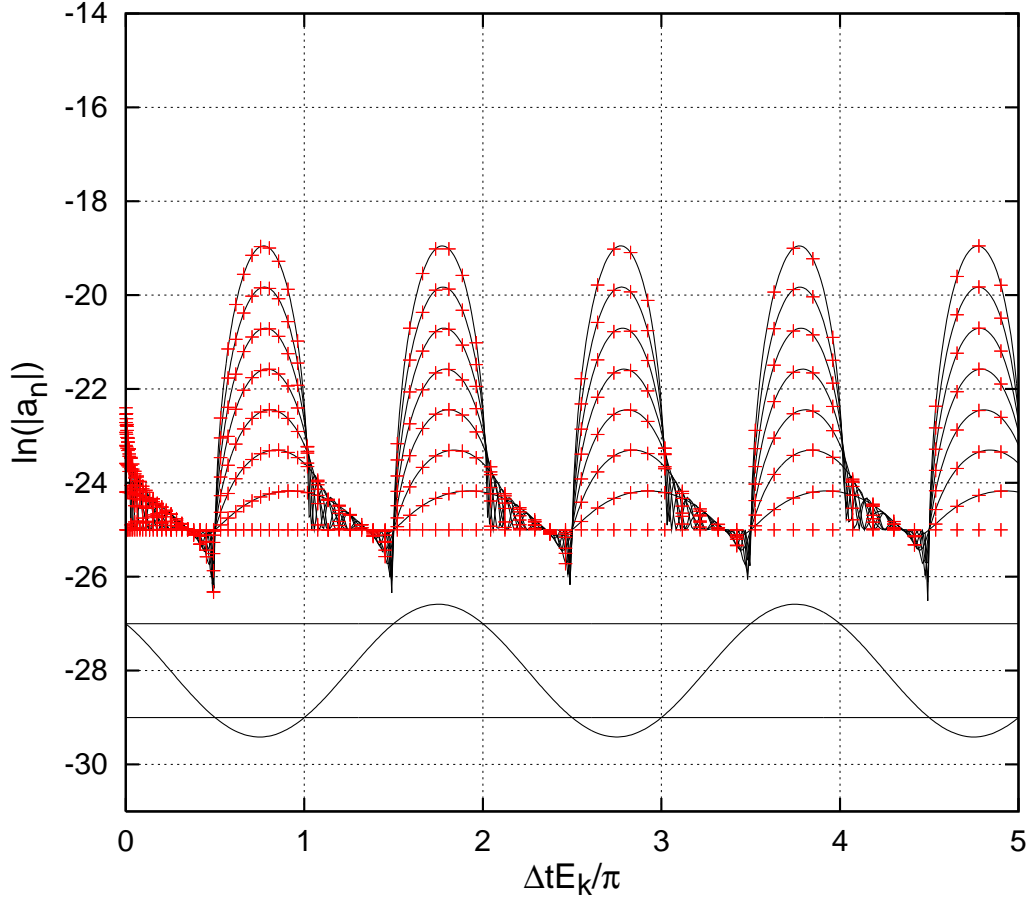}}
	\vspace{0.5truein}
\caption{(Color online) The growth of the error Fourier amplitudes
due to algorithm 2A for seven time steps at $g=5$ and $\dt=0.2$.
The plus signs denotes the algorithm's actual output; the seven
solid lines are the predicted error from the side-band analysis (\ref{eal2a})
for seven time steps.
Centered on -28 is the algorithm's $C$-function for predicting regions
of stability and instability. 
\label{fig3}}
\end{figure}

\newpage
\begin{figure}
	\vspace{0.5truein}
	\centerline{\includegraphics[width=0.8\linewidth]{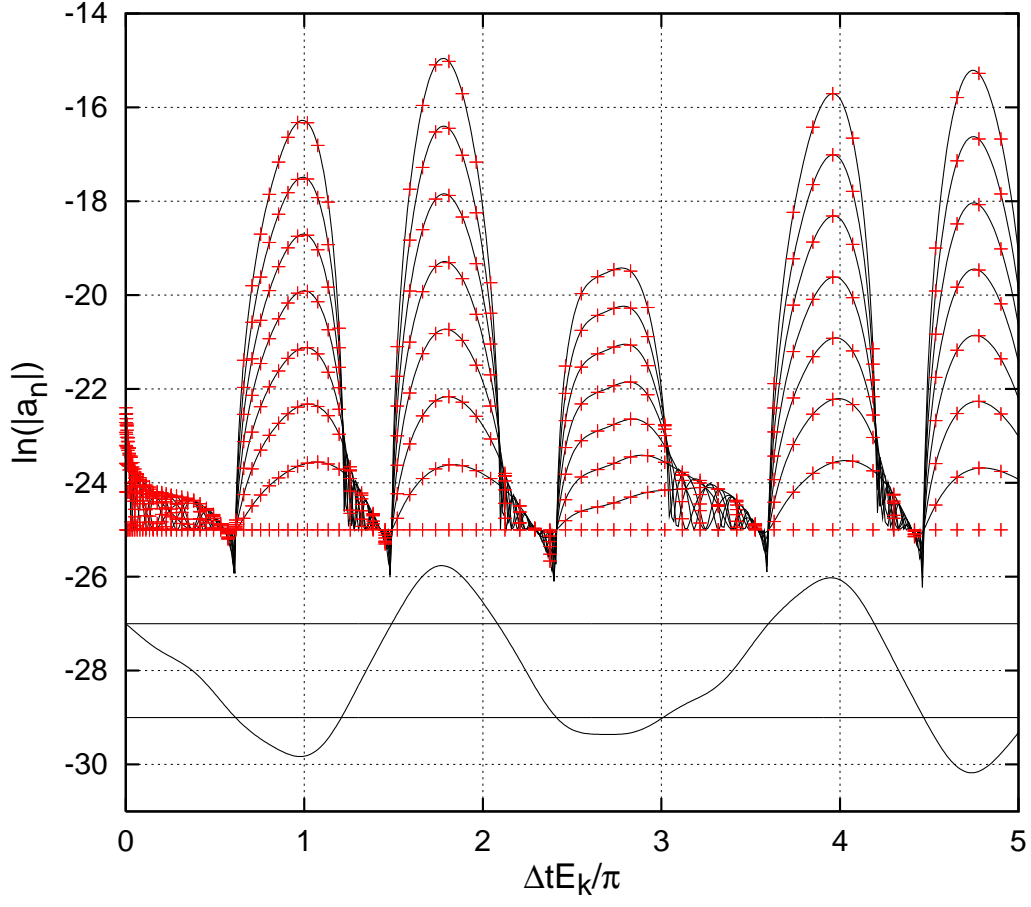}}
	\vspace{0.5truein}
\caption{(Color online) Same as Fig.3 but for
the Forest-Ruth algorithm. The predicted error is given by (\ref{ealfr}). 
\label{fig4}}
\end{figure}

\newpage
\begin{figure}
	\vspace{0.5truein}
	\centerline{\includegraphics[width=0.8\linewidth]{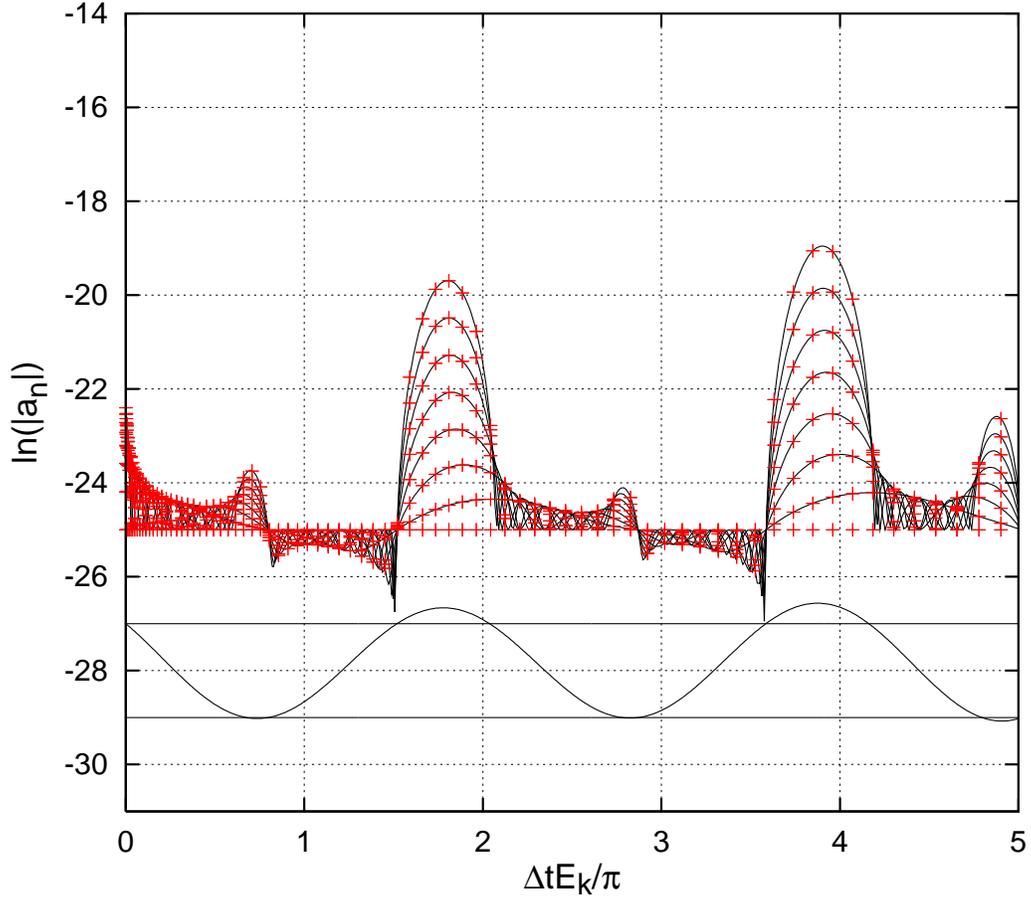}}
	\vspace{0.5truein}
\caption{(Color online) Same as Fig. 3 but for
McLachlan's algorithm. The predicted error is given by (\ref{ealmcl}).
\label{fig5}}
\end{figure}

\newpage
\begin{figure}
	\vspace{0.5truein}
	\centerline{\includegraphics[width=0.8\linewidth]{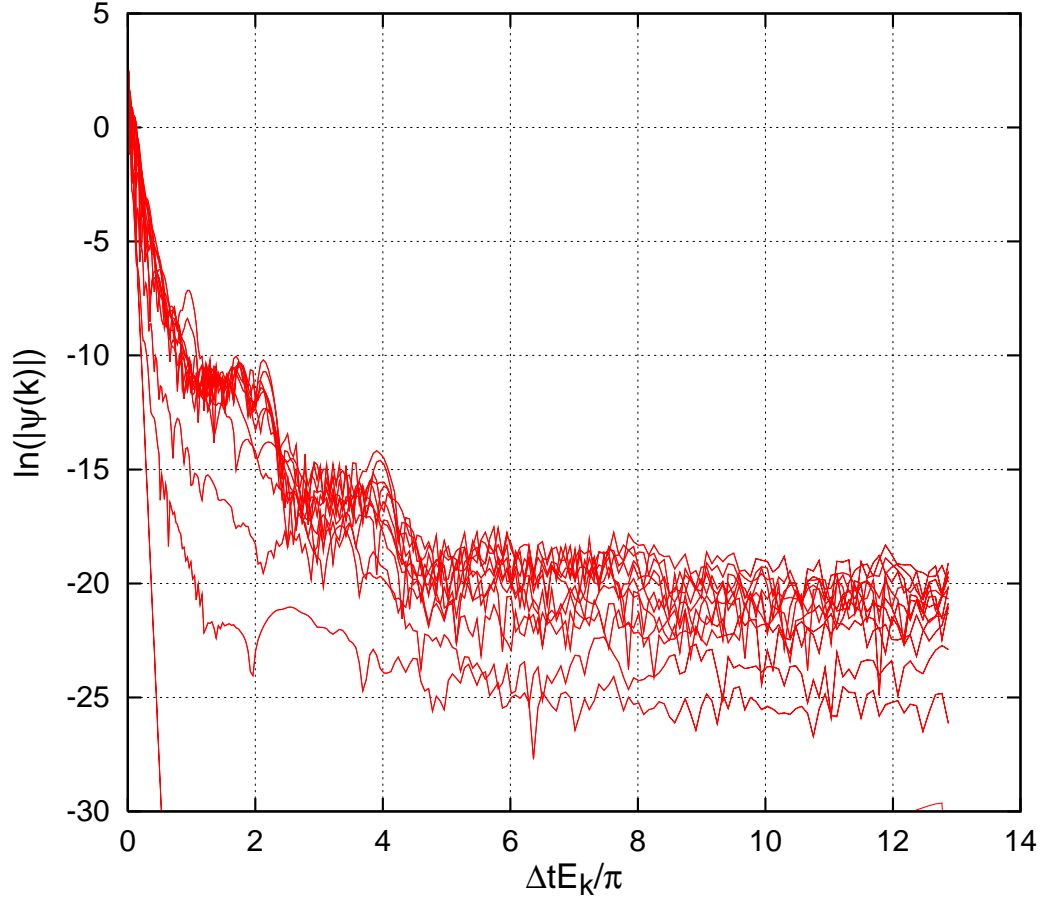}}
	\vspace{0.5truein}
\caption{(Color online) The modulus of the Gross-Pitaevskii momentum wave function
$|\psi(k)|$ at every 100th time-step due to
algorithm 2A. The time step size is $\dt=0.05$.
\label{fig6}}
\end{figure}

\newpage
\begin{figure}
	\vspace{0.5truein}
	\centerline{\includegraphics[width=0.8\linewidth]{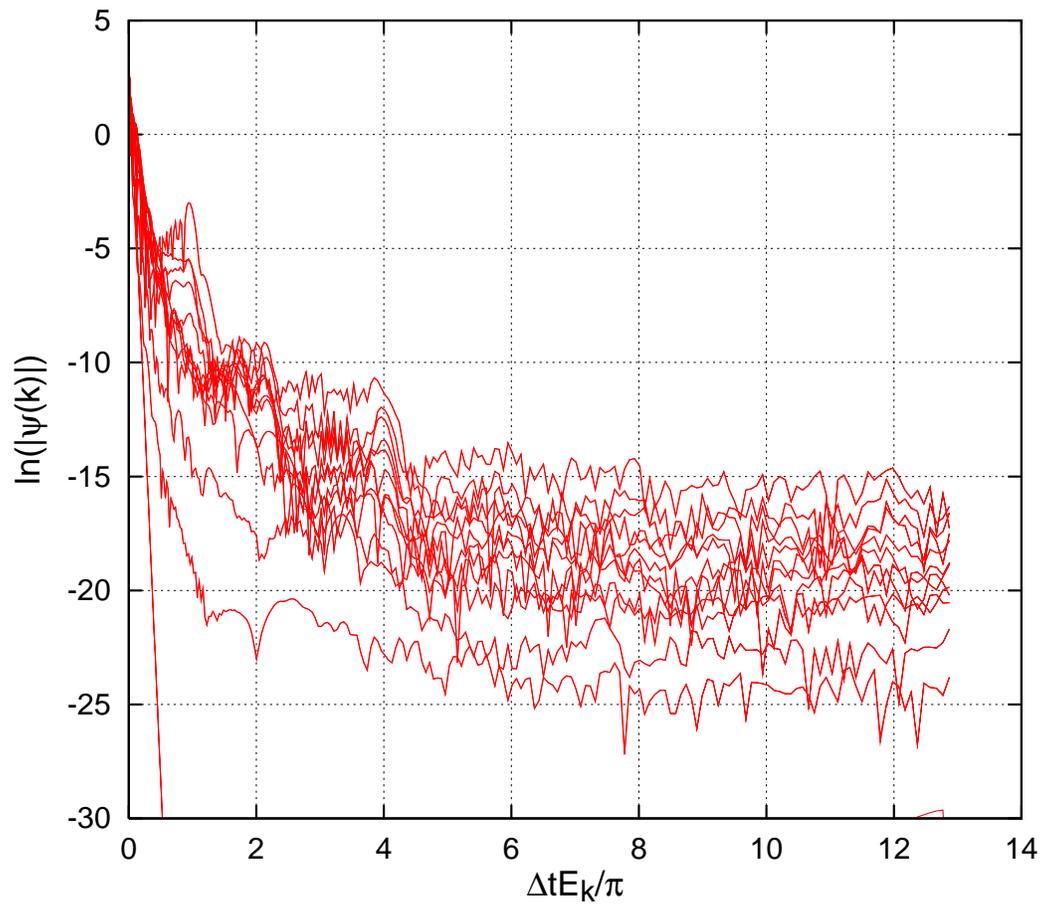}}
	\vspace{0.5truein}
\caption{(Color online) The modulus of the Gross-Pitaevskii momentum wave function
$|\psi(k)|$ at every 100th time-step due to	Forest-Ruth
algorithm. This is the momentum wave function corresponding to the
energy calculation of Fig.1 up to $t=60$.
\label{fig7}}
\end{figure}

\newpage
\begin{figure}
	\vspace{0.5truein}
	\centerline{\includegraphics[width=0.8\linewidth]{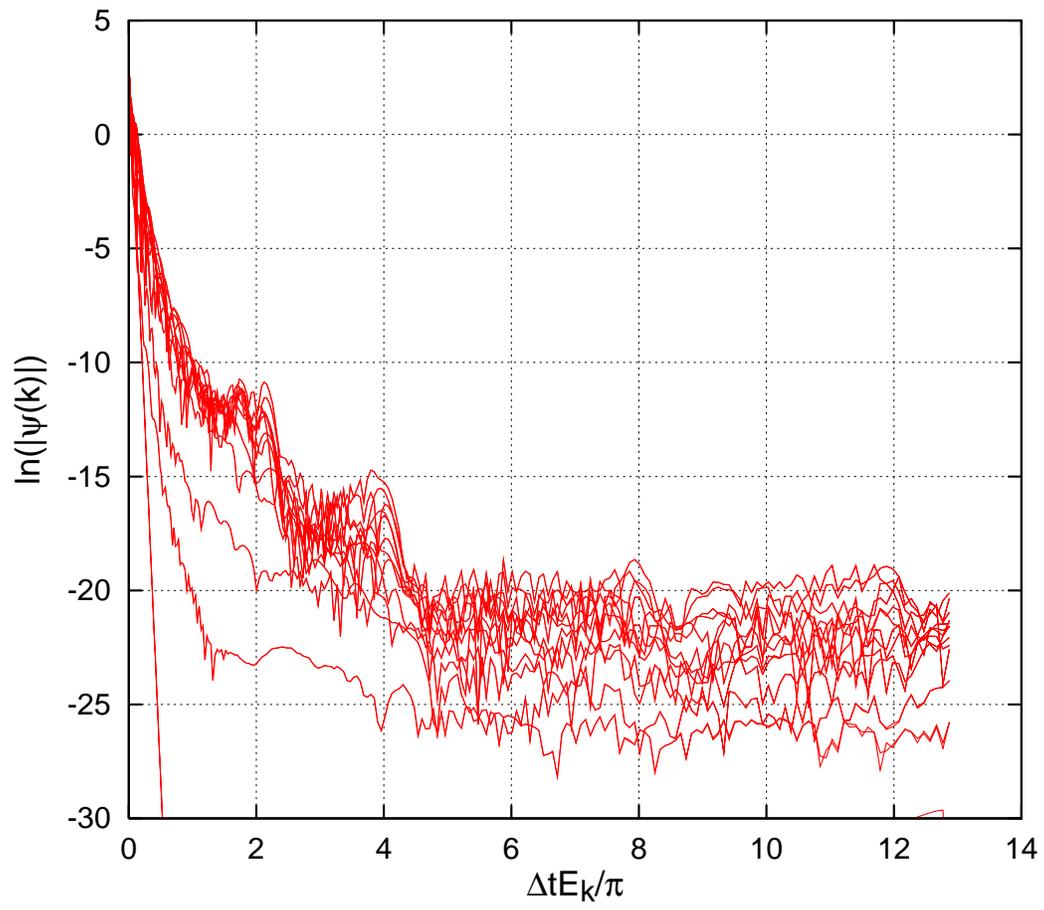}}
	\vspace{0.5truein}
\caption{(Color online) 
Same as Fig.7 for 
McLachlan's algorithm.
\label{fig8}}
\end{figure}


\end{document}